\documentclass{tMOP2e}

\usepackage{epstopdf}
\usepackage{subfigure}
\usepackage{units}

\theoremstyle{plain}

\theoremstyle{definition}

\theoremstyle{remark}

\begin{document}


\articletype{article}

\title{Raman fingerprints on the Bloch sphere of a spinor Bose-Einstein condensate}

\author{
\name{Justin T. Schultz\textsuperscript{a,b}$^{\ast}$\thanks{$^\ast$Corresponding author. Email: j.t.schultz@rochester.edu}, Azure Hansen\textsuperscript{b,c}, Joseph D. Murphree\textsuperscript{b,c}, Maitreyi Jayaseelan\textsuperscript{b,c}, \\ and Nicholas P. Bigelow\textsuperscript{a,b,c}}
\affil{\textsuperscript{a}The Institute of Optics, University of Rochester, Rochester, NY 14627, USA.;
\textsuperscript{b}Center for Coherence and Quantum Optics, University of Rochester, Rochester NY 14627, USA.;
\textsuperscript{c}Department of Physics and Astronomy, University of Rochester, Rochester, NY 14627, USA.}
\received{\today}
}

\maketitle

\begin{abstract}
We explore the geometric interpretation of a diabatic, two-photon Raman process as a rotation on the Bloch sphere for a pseudo-spin-$\nicefrac{1}{2}$ system. The spin state of a spin-$\nicefrac{1}{2}$ quantum system can be described by a point on the surface of the Bloch sphere, and its evolution during a Raman pulse is a trajectory on the sphere determined by properties of the optical beams: the pulse area, the relative intensities and phases, and the relative frequencies. We experimentally demonstrate key features of this model with a $^{87}$Rb spinor Bose-Einstein condensate, which allows us to examine spatially dependent signatures of the Raman beams. The two-photon detuning allows us to precisely control the spin density and imprinted relative phase profiles, as we show with a coreless vortex. With this comprehensive understanding and intuitive geometric interpretation, we use the Raman process to create and tailor as well as study and characterize exotic topological spin textures in spinor BECs. 

\end{abstract}


\section{Introduction}
Since the experimental realization of the first atomic Bose-Einstein condensates (BECs) in 1995, this macroscopic quantum state of matter has been a versatile medium for studying quantum physics \cite{KBDavis1995,HMAnderson1995,CCBradley1995}. Because atoms can have degenerate spin ground states, the wavefunctions are multicomponent vectors (spinors) with many degrees of freedom \cite{HoPRL1998}. The spinor wavefunctions can be sculpted with the use of external magnetic, electric, rf, and optical fields, thereby enabling the creation of analogs from other fields of physics \cite{QuantumSim, ManyBody}. For example, optical Raman interactions have recently been used to create synthetic gauge fields \cite{Gauge1,Gauge2} and spin-orbit coupling \cite{SOC} allowing for the study of the quantum spin hall effect \cite{QSHE} and demonstration of chiral edge states \cite{EdgeStates} in synthetic dimensions \cite{SyntheticDimensions}.  

Magnetic and optical imprinting techniques have created other topological excitations in BECs such as coreless vortices \cite{CVs,Kevin1}, skyrmions \cite{LSLeslie2009,Skyrmion,Skyrmion2}, and both topological \cite{TopoMonopoles} and Dirac monopoles \cite{DiracMonopoles}. Studying the evolution and interaction of these excitations within a BEC will give us insights into the physical processes that underlie phenomena across various fields of physics. Magnetic imprinting techniques rely on controllably sweeping magnetic field zeros through the condensate \cite{MonopoleTheory} and therefore require high stability currents with very low noise. Although this technique creates 3D structures, the variety of possible excitations depends on the limited morphology of the magnetic field profile, and it is not possible to create more than one distinct excitation at a time. 

Optical imprinting methods rely on the relative intensities and phases of optical beams to generate spin and phase distributions within the atomic cloud, and experiments have so far focused on creating 2D spin textures in BECs \cite{WrightSculpting}, although there are several proposals for creating 3D textures \cite{JanneMonopoles,JanneAliceRings}. However, the variety of possible excitations depends only on the availability of optical beams whose phase and intensity profiles are easily controlled by spatial light modulators \cite{SOSLMbooks} or digital micromirror devices \cite{MicroMirrors}. Some of the first coherent control over atomic vapors was exercised via stimulated Raman adiabatic passage (STIRAP) \cite{ShorePRA1995} utilizing Gaussian laser pulses applied in the counter-intuitive order; however, more recent work has show that diabatic square pulses are sufficient to coherently imprint phase structures into BECs \cite{LSLeslieFingerprints}. 

A thorough understanding of the Raman process is crucial for both creating and characterizing exotic topological spin textures in spinor BECs. In this paper, we show that, in many cases, the Raman imprinting technique can be understood geometrically through the Stokes parameters as a rotation on the Bloch sphere, and thereby allows for both sculpting and interrogating spin textures in a spinor BEC. We focus on a simultaneous, diabatic pulse pair that transports population between spin ground states within a single hyperfine manifold and explore the features that arise from both Gaussian and Laguerre-Gaussian Raman beams in both the two-photon resonant ($\delta=0$) and off-resonant ($\delta\neq0$) regimes.  

\section{Three-level $\Lambda$-System as a Pseudo-spin-$\nicefrac{1}{2}$ System}

\begin{figure}
\begin{center}
\includegraphics{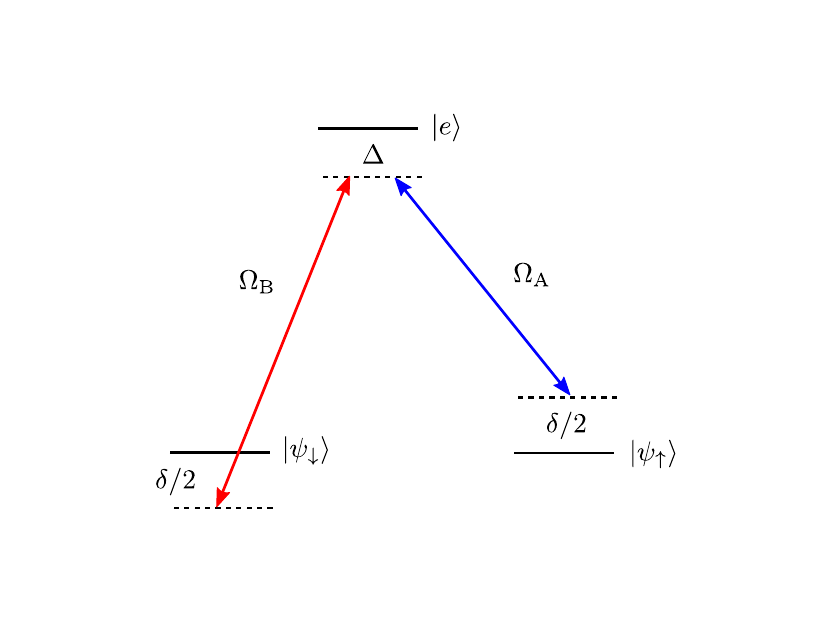}
\caption{Three-level $\Lambda$ system. The states $|\psi_{\uparrow}\rangle$ and $|\psi_{\downarrow}\rangle$ are coupled via two beams with Rabi frequencies $\Omega_{\rm A}$ and $\Omega_{\rm B}$ through $|e\rangle$. The single-photon detuning $\Delta$ is large such that the excited state can be adiabatically eliminated making an effective spin-$\nicefrac{1}{2}$ system.} \label{LambdaSystem}
\end{center}
\end{figure}

To describe the Raman process, we focus on a three-level atom where separate spin ground states $|\psi_{\uparrow}\rangle$ and $|\psi_{\downarrow}\rangle$ are coupled via a third excited state $|e\rangle$ by two optical beams with Rabi frequencies $\Omega_{\rm A}$ and $\Omega_{\rm B}$ as in Figure \ref{LambdaSystem}. The frequencies of the optical beams are tuned such that the single-photon detuning $\Delta$ is large and no population enters the excited state. We can therefore adiabatically eliminate the excited state \cite{AdiabaticElim1,AdiabaticElim2} and describe the atom as a pseudo-spin-$\nicefrac{1}{2}$ system by the two-component spinor ${\bm \psi}=(\psi_{\uparrow},\psi_{\downarrow})^{\rm T}$. 
In optics, another familiar pseudo-spin-$\nicefrac{1}{2}$ system is the transverse polarization of light, which can be described by a point on the surface of the Poincar\'{e} sphere \cite{ArdavanNJP2007}. For any particular transverse polarization state, the coordinates on the Poincar\'{e} sphere that describe it are given by the Stokes parameters \cite{Mueller, Fano-Stokes, Schmieder}. To geometrically describe the spin state of an atomic system, we then begin with the atomic Stokes parameters
\begin{equation}
\begin{array}{cc}
S_0 =& |\psi_{\uparrow}|^2 + |\psi_{\downarrow}|^2\\
S_1 =& 2{\rm Re}\left\{\psi_{\uparrow}^*\psi_{\downarrow} \right\}\\
S_2 =& 2{\rm Im}\left\{\psi_{\uparrow}^*\psi_{\downarrow} \right\}\\
S_3 =& |\psi_{\uparrow}|^2-|\psi_{\downarrow}|^2.
\end{array}
\end{equation}
Because the state vector is normalized, $S_1^2+S_2^2+S_3^2 = 1$ and the state of the atomic system can be represented as a point on the surface of a unit sphere---the Bloch sphere. In this framework, we can view the time evolution of the system during the Raman process as a curve on the surface of the sphere that connects the original state to the final state. 

In the rotating wave approximation \cite{Shore}, the spinor evolves via 
\begin{equation}
\frac{\partial {\bm \psi}}{\partial t} = \frac{i}{4\Delta} \left[\begin{array}{cc}|\Omega_{\rm A}|^2-2\Delta\delta & \Omega_{\rm A}^*\Omega_{\rm B}\\ \Omega_{\rm A}\Omega_{\rm B}^* & |\Omega_{\rm B}|^2 +2\Delta\delta \end{array} \right]{\bm \psi}.
\end{equation}  
For optical pulses that are square in time, the Rabi frequencies are constant over the interaction time, and the system of differential equations can be integrated directly to find ${\bm \psi}(t) = \exp\left(i\frac{\Omega_0 t}{2}\right){\bf M}(\Omega,\alpha,\phi,t){\bm \psi}(t=0)$. The generalized Rabi frequency is given by 
\begin{equation}
\Omega = \sqrt{\Omega_0^2-2\delta\Omega_0\cos 2\alpha_0+\delta^2},
\end{equation}  
where $\Omega_0 = \left(|\Omega_{\rm A}|^2+|\Omega_{\rm B}|^2 \right)/4\Delta$ is the Rabi frequency when the two-photon detuning is zero ($\delta = 0$) and $\tan\alpha_0=|\Omega_{\rm A}|/|\Omega_{\rm B}|$. The Rabi frequencies are complex, $\Omega_j = |\Omega_j|e^{i\phi_j}$, and their relative phase is $\phi = \phi_{\rm A}-\phi_{\rm B}$. The matrix ${\bf M}$ can be written as 
\begin{equation}
{\bf M} = \cos\frac{\Omega t}{2}{\bf I} +i\sin\frac{\Omega t}{2}{\bf P}(\alpha,\phi). 
\end{equation}
Here, ${\bf I}$ is the $2\times 2$ identity matrix, and ${\bf P}$ is a three-dimensional version of the pseudo-rotation matrix \cite{PseudoRotationMatrix}, 
\begin{equation}
{\bf P} = \left(\begin{array}{cc} \cos 2\alpha & \sin 2\alpha e^{-i\phi}\\\sin 2\alpha e^{i\phi} & -\cos 2\alpha \end{array} \right).
\end{equation}
The parameter $\alpha$ describes the relative strengths of the Rabi frequencies, while taking into account the effect of the two-photon detuning ($\delta$), 
\begin{equation}
\tan 2\alpha = \frac{\tan 2\alpha_0}{1-\frac{\delta}{\Omega_0\cos 2\alpha_0} }. 
\end{equation}

\begin{figure}
\begin{center}
\includegraphics{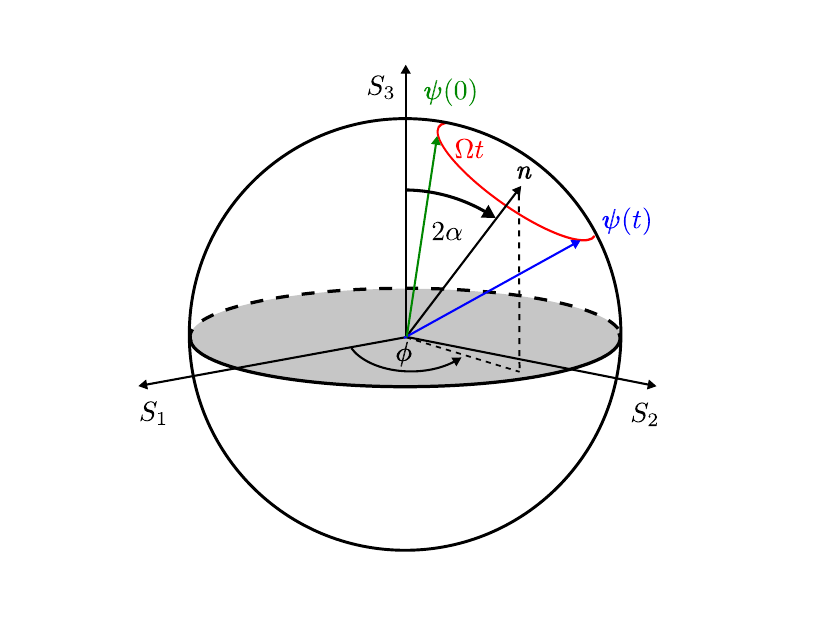}
\caption{The Bloch Sphere. The time evolution of the state ${\bm \psi}(t)$ is a rotation of the initial state ${\bm \psi}(0)$ about the vector ${\bm n}$ by an angle $\Omega t$. The angles $\alpha$ and $\phi$ which describe the orientation of ${\bm n}$ are controlled by the relative intensities and phases of the Raman beams.} \label{BlochSphere}
\end{center}
\end{figure}

The Raman interaction is a rotation of the state on the Bloch sphere \cite{Sakurai} as shown in Figure \ref{BlochSphere}. We can rewrite 
\begin{equation}
{\bf M} = e^{i\frac{\Omega t}{2}{\bm n}\cdot{\bm\sigma}},
\end{equation} 
which is just a rotation by angle $\Omega t$ of the state around the vector 
\begin{equation}
{\bm n} = \left(\begin{array}{c}\sin 2\alpha\cos\phi\\\sin 2\alpha\sin\phi\\\cos 2\alpha \end{array} \right),
\end{equation}
where ${\bm \sigma} = \left({\bm \usigma}_x,{\bm \usigma}_y,{\bm \usigma}_z \right)^{\rm T}$ is the vector of Pauli spin matrices.
With this geometric interpretation, we can explain some of the common features of and gain insight into Raman-coupled pseudo-spin-$\nicefrac{1}{2}$ systems. This description is applicable to any optically addressable three-level system, but the application to a spinor BEC is of particular importance because the spatial extent of the atomic cloud captures the spatial variations in the optical parameters.  Furthermore, Raman processes are being explored both experimentally and theoretically as ways of producing features in BECs such as spin textures, synthetic fields, singularities, and even magnetic monopoles. 

\section{Experimental Setup}

In order to experimentally demonstrate some of the features of the Raman interaction, we use the method which appears in detail in \cite{Kevin1,Kevin3}. We create a spin-polarized $^{87}$Rb BEC of $\sim 6.5\times 10^{6}$ atoms in the $|\psi_{\uparrow}\rangle\equiv |F=2,m_f=2\rangle$ state in a magnetic trap. The atoms are released from the trap and expand for $9\,$ms before two copropagating simultaneous pulses with square temporal profiles couple $|\psi_{\uparrow}\rangle$ to $|\psi_{\downarrow}\rangle\equiv |F=2,m_f=0\rangle$ via $|e\rangle \equiv |F'=1,m_f'=1\rangle$ on the rubidium D$_1$ line. A small bias field of $\sim 11\,$Gauss makes the states individually addressable and defines the quantization axis along which the Raman pulses propagate.

The pulses are derived from the same frequency-locked laser and temporally shaped with acousto-optic modulators (AOMs), which also adjust both the single-photon detuning $\Delta = 440\,$MHz and the two-photon detuning $\delta$. Changing the laser frequency and the frequencies of the AOMs allows control of both $\Delta$ and $\delta$. The widths of the square pulses are typically $1$--$50\,\umu$s. The Raman beams are Gaussian unless sent through a spiral phase plate with an azimuthally varying thickness which makes them Laguerre-Gaussian with donut intensity profiles and azimuthally varying phases. 

After the Raman interaction, an inhomogeneous magnetic field gradient spatially separates the atomic spin states via the Stern-Gerlach effect, and the spatial density distributions are imaged simultaneously with standard absorption imaging. In some instances, before the Stern-Gerlach operation, a second Raman pulse pair with Gaussian transverse profiles is applied $10\,\umu$s after the first pulse pair to interfere the two spin states before they are imaged.  

\section{Two-photon Resonance}

It is illustrative to examine a simplified version of the solution to understand the essential features of the Raman process. Take the case where the system is on two-photon resonance, that is, $\delta =0$. The evolution of the spin state is represented by a circle on the sphere with its center at the point $(\sin 2\alpha_0\cos\phi, \sin 2\alpha_0\sin\phi,\cos 2\alpha_0)$ as can be seen in Figure \ref{BlochSphere} for the case $\Omega = \Omega_0$, $\alpha = \alpha_0$. The state vector rotates around ${\bm n}$ by an angle $\Omega_0 t$. The amplitude and frequency of the Rabi oscillations \cite{Allen} are set by $\alpha_0$ and $\Omega_0$, respectively.

\subsection{Raman Waveplate}
For equal Rabi frequencies, $\alpha_0 = \pi/4$, the vector ${\bm n}$ lies in the $S_1$-$S_2$ plane. If the initial state is in one of the spin eigenstates, then the evolution of the state is represented by a great circle containing the north and south poles. If the Raman beams are also plane waves, the two-photon interaction serves as a Raman waveplate \cite{RamanWaveplate}. The matrix ${\bf M}$ simplifies to the form of a Jones matrix for an arbitrary waveplate \cite{Jones, Jerrard} written in the right-/left-hand circular basis with retardance $\Omega_0 t$ and angle $\phi/2$. 

\begin{figure}
\begin{center}
\includegraphics{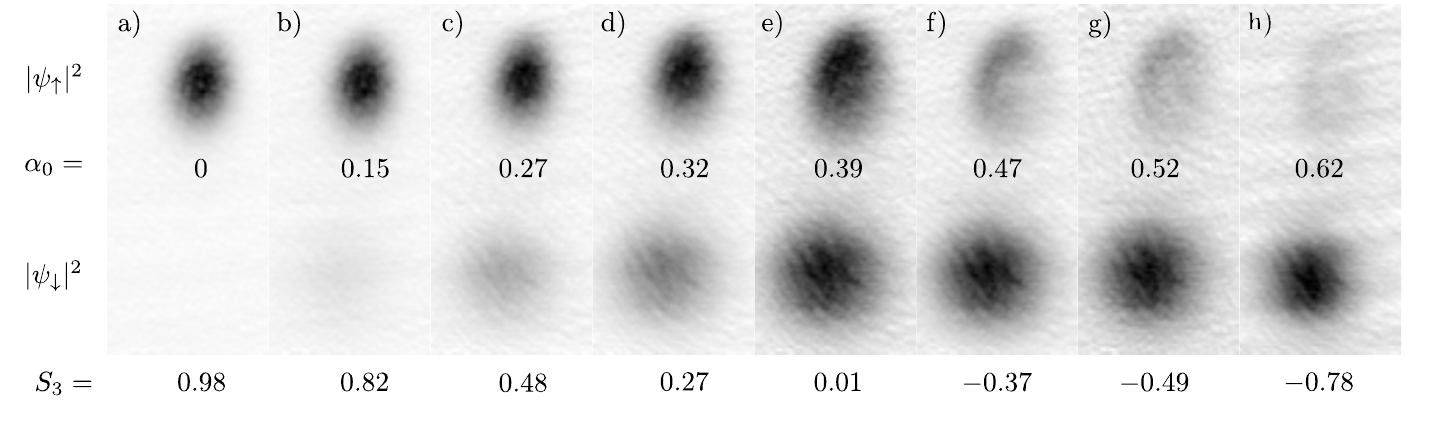}
\caption{Tuning state transfer with $\alpha_0$. These absorption images show the changing ratio of population in each spin state as $\alpha_0$ is varied. For $\Omega_{0} t =\pi$, complete transfer should occur at $\alpha_0 =\pi/4$.} \label{S3vAlphaNOT}
\end{center}
\end{figure}

An optical waveplate can be described as a rotation on the Poincar\'{e} sphere about a vector ${\bm n}$ that lies in the $S_1$-$S_2$ plane. A half-wave plate rotates the state vector by $\pi$ around ${\bm n}$ corresponding to changing linear polarizations (represented by points on the equator) to  linear polarizations and changing elliptical polarizations to elliptical polarizations of the opposite handedness but with the same ellipticity. For circular polarizations (represented by the north and south poles), a half-wave plate has the effect of changing one handedness to the other.  Comparing this to the Raman waveplate, we see that only when the strengths of the Rabi frequencies are equal ($|\Omega_{\rm A}|=|\Omega_{\rm B}|$) is it possible to transfer all the population from one spin state to the other with a $\pi$-pulse ($\Omega_0 t=\pi$). Changing $\alpha_0$ for a fixed $\Omega_0t$ changes the ratio of the population in the two spin states. Figure \ref{S3vAlphaNOT} shows absorption images of the spatially separated spin states after the Raman interaction. As $\alpha_0$ is varied by changing the ratio of the intensities of the Raman beams, the percentage of population transfered (and therefore $S_3$) changes. 

The other common optical waveplate is a quarter-wave plate which rotates the polarization state vector on the Poincar\'{e} sphere by $\pi/2$,  changing circular polarization to linear polarization and changing linear polarization to elliptical or circular (or leaving the polarization linear) depending on the waveplate angle. Therefore, we can use the Raman waveplate not only to generate arbitrary superpositions of the two spin eigenstates but to measure the Stokes parameters of the atomic system \cite{RamanWaveplate}. 

An inhomogeneous magnetic field gradient separates spin eigenstates via the Stern-Gerlach effect, acting analogously to a polarizing beam splitter that separates right- and left-handed circular polarization. In conjunction with absorption imaging, the Stern-Gerlach field gradient allows us to measure the populations of the spin states; however, this technique only shows the squared amplitudes at the north and south poles of the Bloch sphere. The Raman process can overcome this limitation by rotating the squared amplitudes at the antipodal points on any other axis to the north and south poles. 

\begin{figure}
\begin{center}
\includegraphics{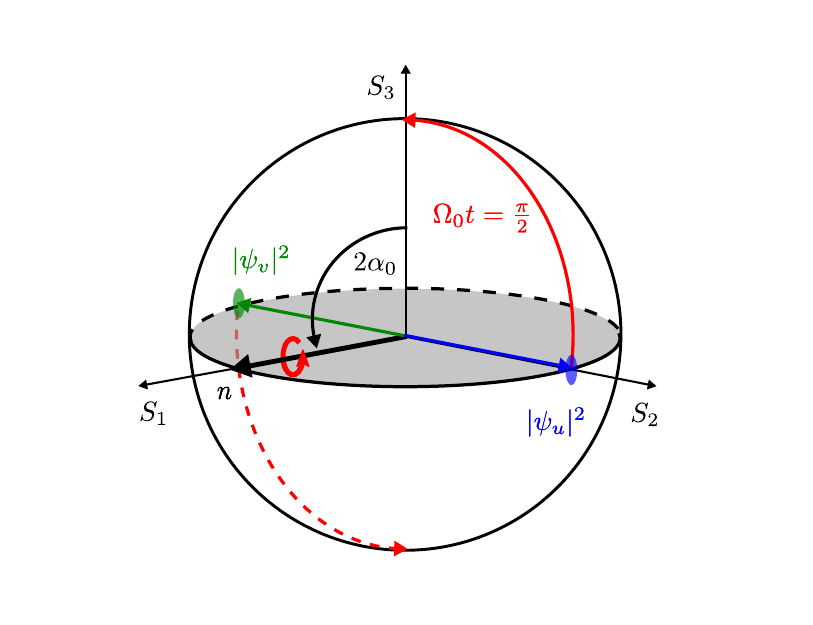}
\caption{Bloch Sphere for a Raman Waveplate. For pair of plane wave pulses with a relative phase $\phi=0$ and equal Rabi strengths $\alpha_0=\pi/4$, the vector ${\bm n}$ lies along the $S_1$-axis. When $\Omega_0t=\pi/2$, the interaction rotates the points corresponding to $|\psi_{u}|^2$ and $|\psi_{v}|^2$ to the north and south poles, respectively. The density profiles can then be separated using a magnetic field gradient and imaged with absorption imaging. The resulting density distributions allow for a measurement of $S_2 = |\psi_{u}|^2-|\psi_{v}|^2$. } \label{RWRotation}
\end{center}
\end{figure}

For example, consider writing the spinor wavefunction in a new orthogonal basis ${\bm \psi} = \left(\psi_{u},\psi_{v}\right)^{\rm T}$ such that $S_2 = |\psi_{u}|^2-|\psi_{v}|^2$. The point on the surface of the Bloch sphere that intersects the positive $S_2$-axis corresponds to all the population being in state $|\psi_{u}\rangle$, and the point where it intersects the negative $S_2$-axis corresponds to all population being in $|\psi_{v}\rangle$. A Raman pulse pair with $\Omega_0 t = \pi/2$, $\alpha_0 = \pi/4$, and $\phi=0$, will rotate the state vector on the Bloch sphere such that $|\psi_{u}|^2$ and $|\psi_{v}|^2$ of the state are represented by the north and south poles, respectively as in Figure \ref{RWRotation}. Therefore, the Raman waveplate allows for a measurement of $S_2$ using Stern-Gerlach state separation and absorption imaging. 

This technique interferes the spin state amplitudes with a relative phase determined by the phase of the Raman laser beams, thereby revealing the inherent phase between the atomic spin components through the interference pattern. Finding the Stokes parameters allows us to reconstruct the spinor wavefunction up to a global phase and will be an important way to characterize the evolution of topological spin textures and other spin systems \cite{Ueda}.

\subsection{Raman Fingerprints}
So far we have been focusing on using the Raman process as a tool to characterize the state (amplitude and relative phase) of a spin system. However, the Raman process is an avenue for creating interesting spin textures in BECs, and examining the resulting atomic system can reveal signatures of the Raman process that created them. Leslie {\it et al}. showed that fringes appear in the population of the spin states due to the transverse intensity gradients of the optical beams and associated light shifts in both a traditional STIRAP and a diabatic pulse pair configuration \cite{LSLeslieFingerprints}. These fringes were described as `fingerprints' of the Raman beams' spatial dependence left on the atomic cloud. Here, we expand the explanation of these features to the Bloch sphere. 

\subsubsection{Gaussian Raman Beams}\label{GaussianRamanBeams}
Simultaneous diabatic pulses require less optical power than STIRAP with Gaussian pulses, and their pulse areas ($\Omega_0 t$) are easier to measure. The high intensity of the Gaussian pulses adds a spatially dependent phase from the AC Stark shift to the atomic cloud creating the concentric rings \cite{Kevin1,LSLeslieFingerprints}. Even for simultaneous, diabatic pulses, with high optical power, the intensity gradient and therefore the gradient of $\Omega_0$ becomes large leading to fringes in the spin states of the atomic clouds. For fixed $\alpha_0$, this can be thought of as a rotation about the same ${\bm n}$ on the Bloch sphere at each point in the BEC; however, because the value of $\Omega_0$ is spatially dependent, areas with high intensity and high $\Omega_0$ may make several full rotations about ${\bm n}$ while portions of the cloud with low intensity make less than one rotation. Leslie, {\it et al}. showed images of the cloud for large $\Omega_0t$; when the gradient of $\Omega_0t$ increases, the Rabi flopping frequency changes more drastically over a shorter distance, causing the concentric rings in the density profile to increase in number and appear closer together, near the extrema of $\nabla\Omega_0t$.  

\subsubsection{Laguerre-Gaussian beams: Coreless Vortices}\label{CVSection}
\begin{figure}
\begin{center}
\includegraphics{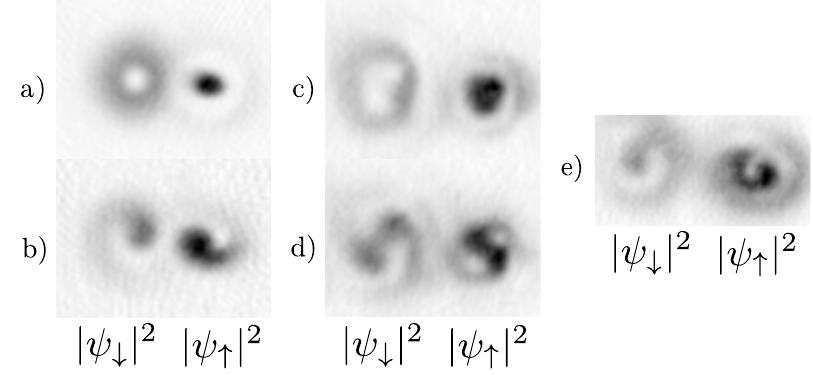}
\caption{Coreless vortices. Absorption images show the Stern-Gerlach-separated density profiles of the spin states. A vortex  in $|\psi_{\downarrow}\rangle$ with (a) $\ell=1$ and  (c) $\ell =-2$ surrounds a non-rotating core in $|\psi_{\uparrow}\rangle$. Atom interference via the application of a Raman waveplate pulse pair confirms the azimuthal phase of (b) $\ell=1$ and (d) $\ell=-2$. The interference pattern (e) shows the radially dependent phase accumulated via the AC Stark shift from the spatially dependent beam intensities. } \label{CV}
\end{center}
\end{figure}

Using the combination of Laguerre-Gaussian and Gaussian modes as the Raman beams creates a coreless vortex: a vortex in one spin state surrounding a non-rotating core in the other spin state (see Figure \ref{CV}). The density profile of the $|\psi_{\downarrow}\rangle$ state looks Laguerre-Gaussian, and the state picks up the difference in phases between the Gaussian and Laguerre-Gaussian beams; that is, the spin state possesses an azimuthally varying phase---a vortex. 

\begin{figure}
\begin{center}
\includegraphics{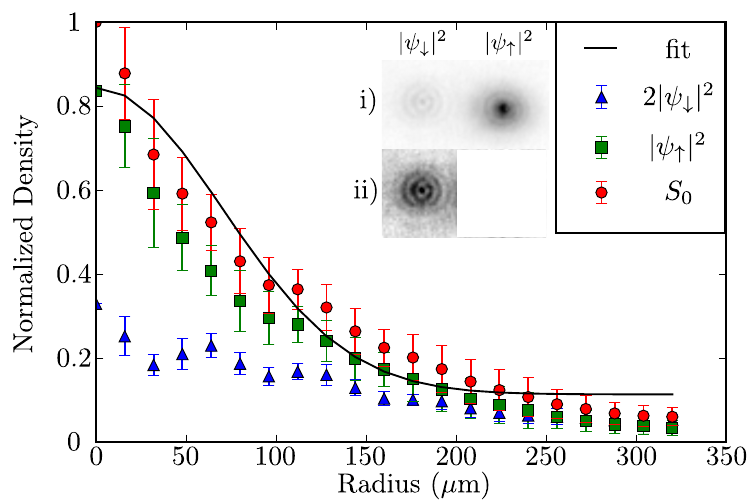}
\caption{Incomplete state transfer. The radial profiles of the spin states show that for a spatially dependent beam, it can be difficult to meet all requirements for complete transfer between states. In the plot, $|\psi_{\downarrow}|^2$ is doubled to emphasize the concentric rings which can be seen in the absorption images (inset) for (i) both spin states and a (ii) re-scaled image of $|\psi_{\downarrow}|^2$. }\label{PartialTrans} 
\end{center}
\end{figure}

Beams with different spatial profiles cause $\alpha_0$ as well as $\Omega_0$ to be spatially dependent. This complicates the structures created in the spin states, because it is possible that with the combinations of $\alpha_0$ and $\Omega_0$ that populations can be transfered completely from one state to the other only at specific locations, or no location at all as shown in Figure \ref{PartialTrans}. Here, the radial profiles of density distributions of $|\psi_{\uparrow}|^2$ and $|\psi_{\downarrow}|^2$ show that there is never complete transfer from $|\psi_{\uparrow}\rangle$ to $|\psi_{\downarrow}\rangle$. 

\begin{figure}
\begin{center}
\includegraphics{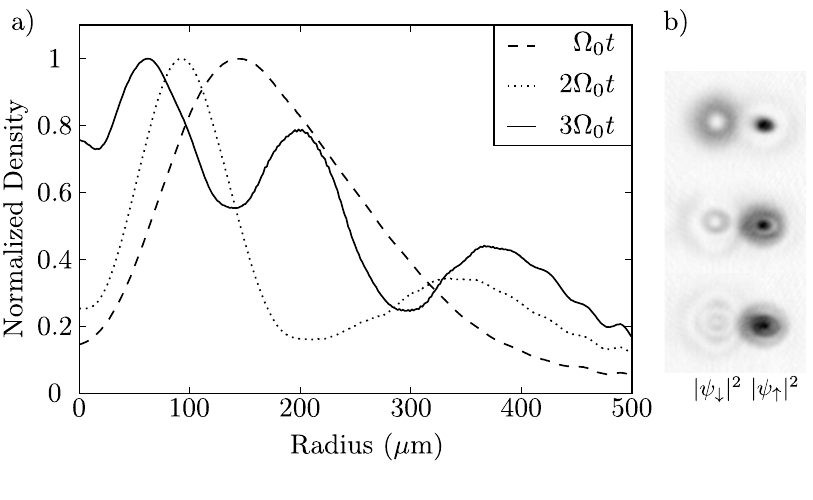}
\caption{Raman fingerprints. As the pulse area ($\Omega_0t$) increases, $\nabla\Omega_0t$ increases leading to a rapid change in the Rabi oscillation frequency near extrema. The radial density profiles (a) and absorption images (b) show the increasing spatial Rabi oscillation frequency with increasing pulse area.}\label{GGRingsVOmt} 
\end{center}
\end{figure}

\begin{figure}
\begin{center}
\includegraphics{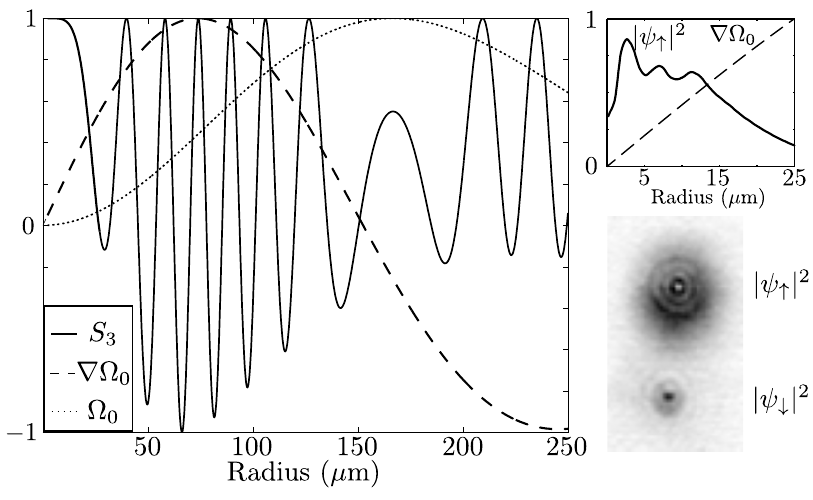}
\caption{Rings near the extrema of $\nabla\Omega_0t$. The radial gradient of $\Omega_0t$ determined from a fit of the Raman beams is plotted with $\Omega_0t$ and the expected radial profile of $S_3$ for large value of $\Omega_0t$. Both the amplitudes of $\Omega_0t$ and $\nabla\Omega_0t$ are scaled for ease of comparison with $S_3$. Concentric rings in the density profiles are concentrated near the extrema of $\nabla\Omega_0t$ rather than at the maximum of $\Omega_0t$, where the Rabi frequency is largest. When the BEC is small compared to the Raman beams, $\nabla\Omega_0t$ is approximately linear, and the concentric rings have a higher visibility near the center of the atomic cloud as seen in a graph of the radial profile of $|\psi_{\uparrow}|^2$ at the time of the Raman interaction (right, top) and the corresponding absorption image after expansion (right, bottom).} \label{GradMax}
\end{center}
\end{figure}

Just as for the Gaussian beams case in \ref{GaussianRamanBeams}, increasing powers of the Raman beams increases the gradient of $\Omega_0t$ thereby increasing the number of concentric vortex rings (Figure \ref{GGRingsVOmt}) which are concentrated near the extrema of the gradient of $\Omega_0t$. In Figure \ref{GradMax} we have plotted the radial gradient of $\Omega_0t$ calculated from fitted Raman beam intensity profiles along with the expected value of $S_3$. The radial density profile of concentric vortex rings in a BEC is plotted with $\nabla\Omega_0t$ in Figure \ref{GradMax} (right). Because the extent of the atomic cloud is small at the time of the interaction compared to the Raman beams, the visibility of the rings is only appreciable near the maximum of the density of the cloud.

An advantage of using a Laguerre-Gaussian Raman beam is the azimuthal variation of the phase. The relative phase of the two spin components is given by the azimuthal coordinate on the Bloch sphere which depends on the Raman beam parameters as $\varphi = \phi- \arctan\left(\tan\left(\Omega_0 t/2\right)\cos 2\alpha_0\right)$. The spatial dependence of $\Omega_0t$ leads to a spatially dependent AC Stark shift \cite{WrightSculpting}, and the azimuthal phase of the vortex allows us to visualize this spatially dependent relative phase through atom interference as seen in Figure \ref{CV}. The interference patterns are not perfect lobes because of this additional radial phase from the Laguerre-Gaussian beam.  The concentric vortices can therefore have different relative phases depending on the relative local intensities at that radius. By interfering concentric vortices with a Raman waveplate pulse, we can get images of interference patterns of the rings and see that the radial interference patterns are offset azimuthally as in Figure \ref{CV}(e).  This provides a signature of the Raman beams not only through the number of rings which is related to the overall power of the beams but also the relative phases which is related to the overall power and $\alpha_0$.    

\section{Tuning the Raman Interaction with $\udelta$}

The Raman process can create interesting spin textures with precise spatially dependent control over the amplitudes and relative phase of the spin components. We can use the two-photon detuning $\delta$ to fine-tune the spatial and relative phase profiles of the created textures. For a non-zero two-photon detuning, $\Omega_0\rightarrow\Omega(\delta)$ and $\alpha_0\rightarrow\alpha(\delta)$, therefore $\delta$ is a parameter that gives us precise control over the imprinted spin textures created in the lab. Changing $\delta$ allows us to change not only the size but also the phase of imprinted features. 

\begin{figure}
\begin{center}
\includegraphics{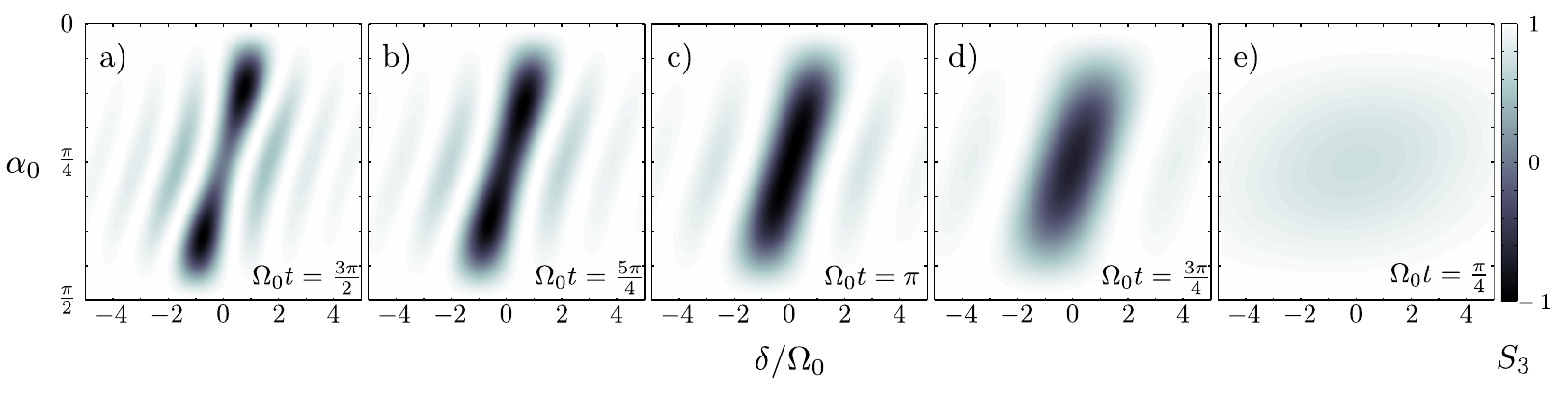}
\caption{Maps of $S_3$ for non-ideal $\alpha_0$ and $\Omega_0t$. For plane waves with $\Omega_0t>\pi$, the resonance splits into two regions where $\alpha_0$ and $\delta$ conspire to allow significant transfer to $|\psi_{\downarrow}\rangle$. Here we show (a) $\Omega_0t =3\pi/2$, (b) $\Omega_0t=5\pi/4$ in comparison to (c) $\Omega_0t=\pi$. For $\Omega_0t<\pi$ such as (d) $\Omega_0t=3\pi/4$ and (e) $\Omega_0t=\pi/4$, $\delta$ and $\Omega_0t$ cannot completely compensate for an insufficient Rabi frequency, meaning complete transfer is no longer possible. } \label{S3vDelta}
\end{center}
\end{figure}

\begin{figure}
\begin{center}
\includegraphics{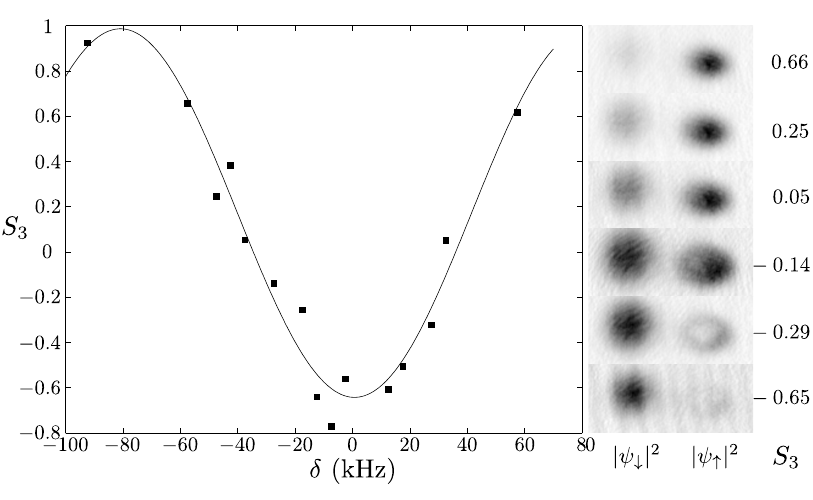}
\caption{Tuning population transfer with $\delta$. $S_3$ is plotted (left) as $\delta$ is varied for $\Omega_0t = 2\pi/5$ and $\alpha_0 \approx \pi/4$. A selection of absorption images (right) from the plot show the change in population transfer with $\delta$. } \label{GG2Photon}
\end{center}
\end{figure}

One major effect of a nonzero $\delta$ is to shift the value of $\alpha$ relative to $\alpha_0$; for $\delta<0$, $\alpha<\alpha_0$, and for $\delta>0$, $\alpha>\alpha_0$. Because $\delta$ modifies $\alpha$, it also modifies the conditions needed to get complete transfer from one spin state to the other. For $\alpha_0\neq\pi/4$, the correct tuning of $\delta$ can result in complete transfer between the states as illustrated theoretically in Figure \ref{S3vDelta} for the case of plane waves. Tuning $\delta$ is a way to narrow the search for the correct parameters for complete transfer or any particular desired percentage of transfer. In Figure \ref{GG2Photon} a graph of $S_3$ as a function of $\delta$ reveals the two-photon resonance condition which increases transfer between the two spin states for a non-ideal $\alpha_0$ and $\Omega_0t$. 

\begin{figure}
\begin{center}
\includegraphics{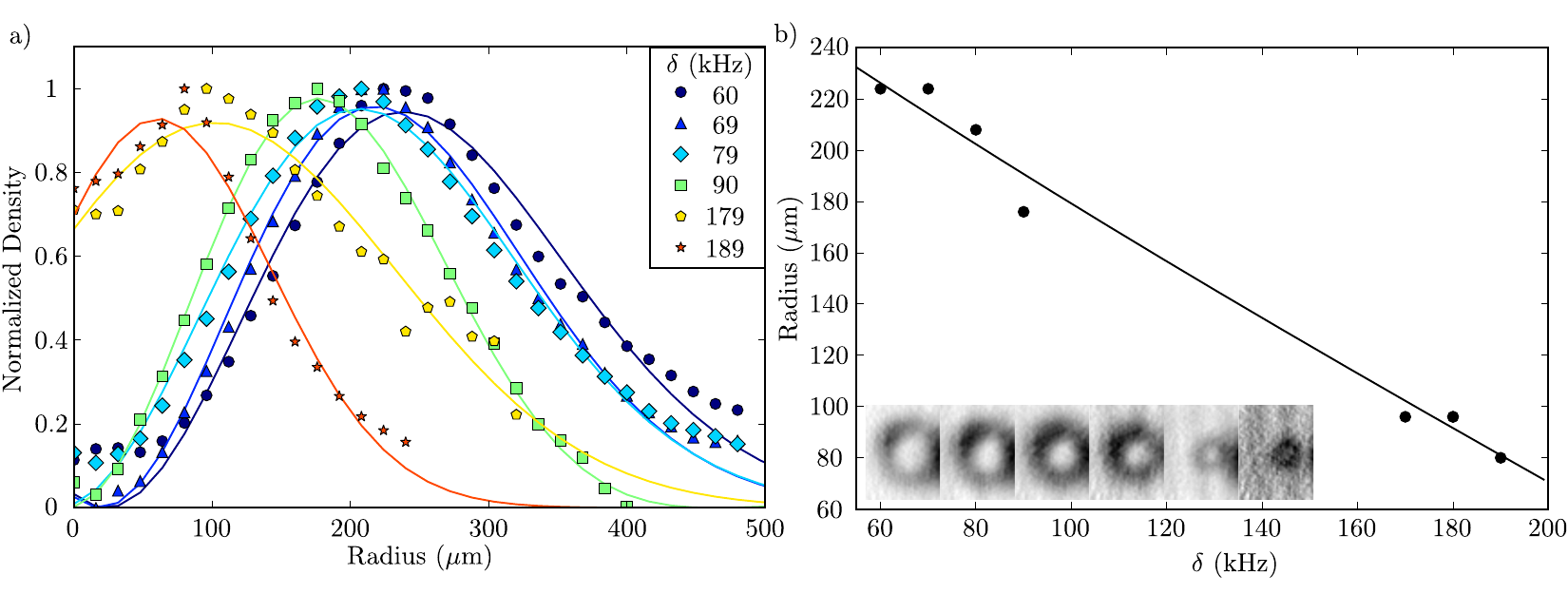}
\caption{Tuning the vortex size. The radial density profiles of the vortex states for various two-photon detunings (a) show that the radius of the vortex can be tuned with $\delta$, with (b) the dependence of the vortex size on $\delta$. Absorption images (b, inset) show the change in radius of the vortices with $\delta$. For larger vortices, the Stern-Gerlach operation did not completely separate the spin states, and the increased density on the right side is from $|\psi_{\uparrow}|^2$.} \label{RingsVdelta}
\end{center}
\end{figure}

\begin{figure}
\begin{center}
\includegraphics{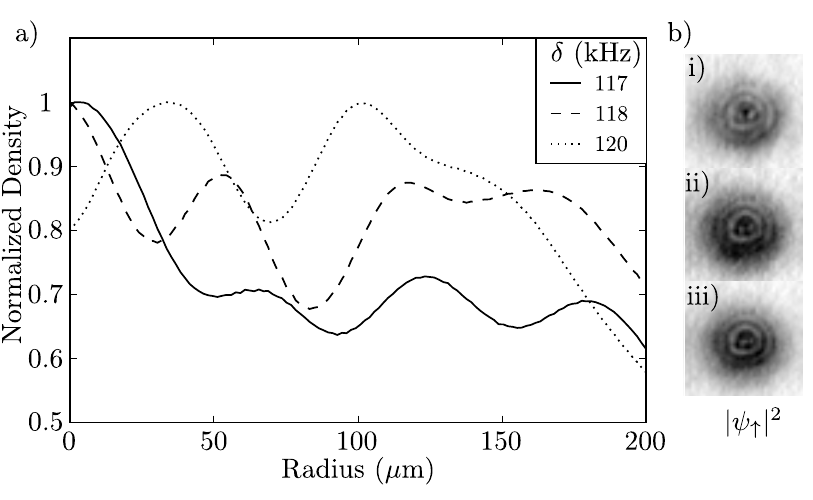}
\caption{Tuning the vortex rings. Here we show radial density profiles (a) and absorption images (b) of concentric vortex rings as $\delta$ is tuned to (i) 117\,kHz, (ii) 118\,kHz, and (iii) 120\,kHz. The two-photon detuning makes small adjustments to the locations of the maxima of the rings, but because the rings are the result of several Rabi oscillations during the Raman pulse, they stay near the maximum of $\nabla\Omega t$.} \label{CrazyRingsVdelta}
\end{center}
\end{figure}

\begin{figure}
\begin{center}
\includegraphics{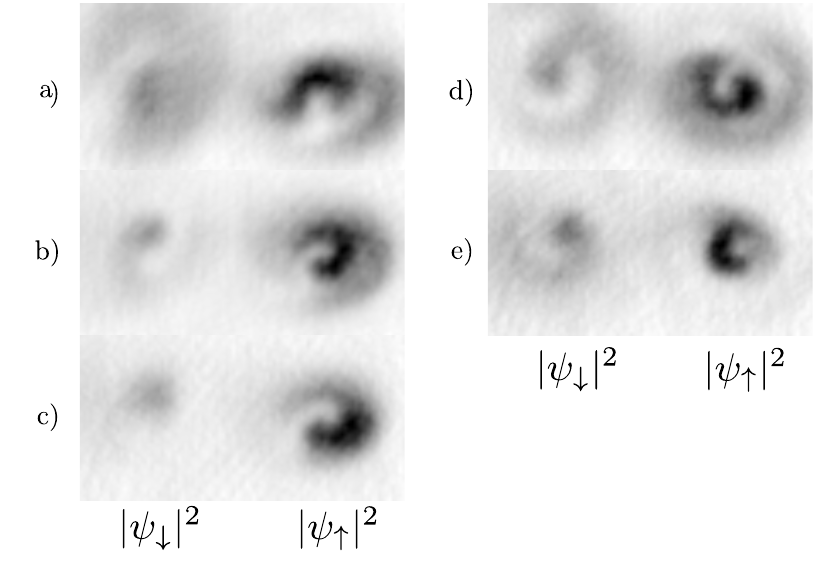}
\caption{Tuning the relative phase with $\udelta$. Changing the two-photon detuning ((a)95\,kHz, (b) 110\,kHz, (c) 115\,kHz) of the Raman beams that create the coreless vortex changes the orientation of the interference pattern in the absorption image and changing the two-photon detuning of the Raman waveplate beams ((d) 110\,kHz, (e) 115\,kHz) that create the atomic interference causes the interference pattern to rotate. In both cases the relative phase introduced between the spin states is dependent on $\delta$.} \label{PhaseVdelta}
\end{center}
\end{figure}

In creating the coreless vortices in \ref{CVSection}, we created a spatially dependent $\alpha_0$, where the maximum population in the $|\psi_{\downarrow}\rangle$ state is achieved when $\alpha_0=\pi/4$ and $\Omega_0t=\pi$. For  $\delta/\Omega_0 <<1$, $\Omega\approx\Omega_0$, and $\delta$ tunes only the value of $\alpha$. Tuning $\alpha$ with the two-photon detuning allows us to change the location of maximum population transfer, thereby changing the size of the imprinted vortex. Figure \ref{RingsVdelta} shows radial profiles and numerical fits of the vortex component of a coreless vortex as the two-photon detuning is changed. For small $\delta$ and $\alpha_0\neq\pi/4$, the maximum of the ring occurs when $\alpha=\pi/4$, that is, when $r_0 \approx \tan\left(\arccos\left(\delta/2\Omega_0\right)\right)$, for a vortex with winding number $|\ell|=1$. 
The sizes of the concentric vortex rings can be changed in a similar manner as shown in Figure \ref{CrazyRingsVdelta}. The change in position is less noticeable since the rings are the result of several Rabi oscillations within the Raman pulse and are therefore more likely to be located near the extrema of $\nabla\Omega t$. 
Similarly, the relative phase between the spin components depends on $\alpha$ via $\varphi = \phi-\arctan\left(\tan\left(\Omega t/2\right)\cos 2\alpha\right)$, implying that it also depends on $\delta$. We can therefore change the relative phase of the spin components by changing the two-photon frequency. In Figure \ref{PhaseVdelta}, a Raman process creates a coreless vortex and then a Raman waveplate pulse interferes the spin components before state separation and imaging. Changing $\delta$ for the beams creating the coreless vortex also changes the orientation of the atomic interference pattern because of this extra phase. Similarly, changing the relative frequencies of the Raman waveplate beams also causes the interference pattern to rotate. The original Raman waveplate scheme relied on changing path lengths in an interferometer with a piezo-electric transducer \cite{RamanWaveplate}; however, the next generation Raman waveplate could rely on changing the relative frequencies between beams since it is easy to control frequencies with high precision.   

\section{Summary and Conclusion}
In conclusion, we have shown that the two-photon Raman process can be interpreted geometrically as a rotation on the Bloch sphere for the state of a pseudo-spin-$\nicefrac{1}{2}$ system. Although this geometric interpretation is applicable to any optically addressable three-level system, a spinor BEC provides a medium for experimentally verifying this interpretation as well as investigating the spatially dependent features that arise from the transverse spatial intensity and relative phase profiles of the Raman beams, which leave fingerprints on the atomic density and phase profiles. Through this description, we gain intuition in how to use the Raman process to both characterize (via the Stokes parameters and a Raman waveplate) and create spin textures. The two-photon detuning $\delta$ provides a fine-tuning parameter allowing us to control the size of the imprinted spin texture and relative phase of the spin states, as we demonstrate on a coreless vortex. 

The diversity of possible spin textures is limited only by the available range of the intensity and phase profiles of the Raman beams \cite{FPBeam,FPBeam2,Kumar}, making this technique important for engineering spin textures analogs of other physical systems and studying their interactions and evolution. For example, the Raman scheme can be used to create and study angular spin-orbit coupling in a spinor BEC with complex optical beams \cite{AngularSOC1, AngularSOC2, AngularSOC3}. Because atoms have multidimensional spin manifolds, there is more to explore in describing more complex Raman interactions as rotations on higher dimensional Bloch spheres or combinations of Bloch spheres \cite{Milione}. This opens a path for using a spinor BEC as a medium for non-orthogonal quantum measurements \cite{SFA} with structured light \cite{SFA2} as well as for using  spin textures in BECs as topological qubits for quantum computing schemes \cite{QuantumComputing1,QuantumComputing2}.

\section*{Acknowledgments}
The authors gratefully acknowledge support from NSF and NASA.

\section*{Disclosure statement}
No potential conflict of interest was reported by the authors.

\end{document}